\documentclass[amsmath,amssymb,twocolumn]{revtex4}
\usepackage{graphicx}
\usepackage{txfonts}
\usepackage{float}
\usepackage{mymacro}
\usepackage{multirow}
\usepackage{color}
\usepackage{bm}
\usepackage{enumerate}

\begin{document}
\title{Experimental demonstration of quantum leader election in linear optics}%
\author{Yuta Okubo$^{1,2}$, Xiang-Bin Wang$^{2}$, Yun-Kun Jiang$^{2}$, Seiichiro Tani$^{2}$, and Akihisa Tomita$^{2}$}
\affiliation{$^1$Department of Frontier Science, University of Tsukuba\\
1-1-1, Tennodai, Tsukuba-shi, Ibaraki 305-8501, Japan}
\affiliation{$^2$ERATO-SORST Quantum Computation and Information Project, JST\\
5-28-3, Hongo, Bunkyo-ku, Tokyo 113-0033, Japan}
\begin{abstract}
Linear optics is a promising candidate to enable the construction of quantum computers. A number of quantum  protocols gates  based on linear optics have been demonstrated. However, it is well-known that these gates are non-deterministic and that higher order nonlinearity is necessary for deterministic operations. We found the quantum leader election protocol (QLE) can be operated deterministically only with linear optics, and we  have demonstrated the nearly deterministic operation which overcomes classical limit. 
\end{abstract}
\maketitle
\section{Introduction}
Linear optics is a useful tool for constructing quantum circuits by utilizing optical interference \cite{KLM}. A number of quantum gates \cite{KYI,zei} were demonstrated with linear optics. 
However, the operation of these gates has been non-deterministic, because deterministic operation requires nonlinear interactions between photons. The success probability of the linear optical CNOT gate is at most 1/4 \cite{CNOT1}, for instance.
Therefore, it is quite difficult to implement large-scale quantum computations with linear optics such as Shor's factorization algorithm\cite{shor}, because the more quantum gates the quantum algorithm requires, the less the success probability.
It is becoming more important to implement deterministic quantum operations.
We have found that the quantum leader election (QLE) protocol can be operated deterministically only with linear optics, and have demonstrated nearly deterministic operation in an experiment with two-photon interference.
\section{Quantum leader election protocol}
Before considering the QLE protocol, let us briefly introduce distributed computing.
A distributed computing system is a set of computers connected via network channels and it offers the fault-tolerance and the scalability of the system.
The system operates autonomously with a distributed algorithm\cite{lyn} that effectively coordinates the computers. 
In a lot of distributed algorithms, computers generate a $token$, which is a kind of script that prescribes the operation or gives permission for them to access shared data.
A computer does some operations based on the token and relays it to the next computer. Then, the next computer receives the
token and does operations, and so on.
After going round the whole network, one distributed algorithm is completed.
If the token is lost during the operation, a unique computer must be specified to restore the only one token, otherwise two or more tokens that tangles the operation may be generated.

Such a unique computer is called the leader, and the decision of the leader is known as the leader election problem \cite{lyn}. Since  only one leader should be chosen before performing any distributed algorithms, the leader election problem is one of the most fundamental problems in the distributed computing. Many studies on the leader election problem have been done in the long history of computer science. We can summarize the leader election problem as :
\begin{description}
\item [Leader election problem]\cite{lyn}\ \\
{\it
Let N be the number of computers connected to the network. Each computer i $\in\left\{1,2,\dots,n\right\}$ has a state
$x_i$ and is initially set to identical values $x_i=c_i$. Set the unique computer $k\in\left\{1,2,\dots,n\right\}$, to $x_k=1$, otherwise, $x_j=0$
$\bigl(j\neq k, j\in\left\{1,2,\dots,n\right\}\bigr).$}
\end{description}
In this problem, there are no third persons such as human beings to determine the leader; therefore, the computers must do the same operations during the protocol. If the network is non-anonymous, ($c_i \not\equiv c_j$ for $i\neq j$), in which all computers have an identifier (i.e., MAC address), this protocol can be solved quite easily by comparing each identifier. The computer that has the largest value of identifier can be elected to the leader.

However if the network is anonymous ($c_i\equiv c_0 \forall i \in \{1,\dots N\}$), in which the computers have no identifiers and they are in an identical initial state, the leader election is no longer trivial. They have to distinguish themselves and finally obtain different identifiers to distinguish one another using the same operation.
In other words, they have to obtain asymmetric states from the symmetric state after identical operations.
It is widely recognized  \cite{ang,lyn,yma1} that no classical algorithm can offer a deterministic solution within a finite time to this problem. The possible classical algorithms are stochastic \cite{ita}, where each computer automatically generates an identifier at random.  Then, some computers generate the same identifiers with a certain probability, and leader election fails.
In contrast, Tani $et\ al$ proposed a new quantum algorithm \cite{tani1} that could solve this problem deterministically within a finite time. The process for the quantum protocol is as follows: 
\begin{description}
\item [Quantum leader election protocol]\ \\
Computers with identical quantum states ($\ket{x_i}\equiv\ket{c_0} \forall \{1\dots N\}$) are connected via quantum and classical communication channels. A unique leader is elected after all computers repeats the following processes for N-1 times for N parties:
\begin{enumerate}
\item Entangled state generation,
\item Quantum communication,
\item Local quantum operation, and
\item Measurement and classical communication.
\end{enumerate}
\end{description}
The quantum state of each computer is identical in this protocol. After quantum communication, the computers build a quantum correlation by sharing entanglement. They can finally derive a non-identical measurement result.

We have demonstrated the two-party QLE protocol, which makes the implementation simple, because it only requires one execution of processes (I) $\sim$ (IV) without any ancillary qubits.
\subsection*{Implementation of two-party QLE}
\begin{figure}[t]
\begin{center}
\includegraphics[width=8.5cm]{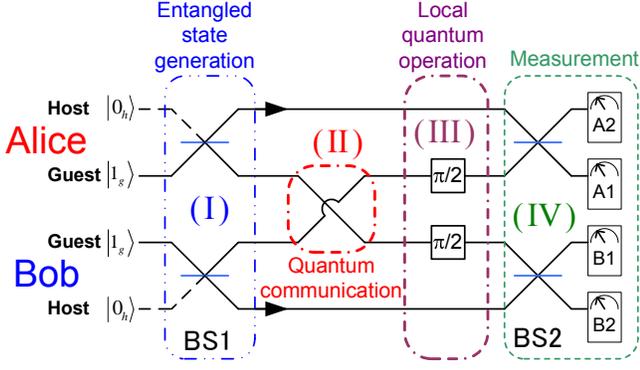}
\caption{(Color online) Implementation of two-party quantum leader election with linear optics: (I) Entangled state generation, (II) Quantum communication channel, (III) Local quantum operation, and (IV) Measurement. Both Alice and Bob have two ports: $host$ and $guest$}\label{cir}
\end{center}
\end{figure}%
There is a schematic of the two-party QLE circuit is shown in FIG.\ref{cir}. 
Since all parties must perform exactly the same operations, Alice's setup is line symmetric to Bob's.
Both Alice and Bob have two optical ports defined as the $host$ and the $guest$, respectively, and the four detectors A1, A2, B1, and B2 are named according to the symmetry shown in FIG.\ref{cir}
Here, the qubit is represented by the photon number, where single photon state $\ket{1}$ represents qubit 1 and vacuum state $\ket{0}$ represents a qubit 0.
The circuit consists of four parts (I) $\sim$ (IV), carrying out the QLE process are given as follows:

\paragraph{Entangled state generation}\ \\
\ \ \ Alice and Bob must prepare identical symmetric state $\ket{\psi_0}\equiv\ket{0_h,1_g}_\text{A}\otimes\ket{0_h,1_g}_\text{B}$, which corresponds to a single photon pair emerged in their guest ports. 
Note that photons must be indistinguishable with each other, otherwise we would contravene the identity of initial state.
Then single photon entangled states are generated at Alice and Bob by the 50:50 beam splitters (BS1) as shown in FIG.\ref{cir}.
\begin{align}
\ket{\psi_0}\equiv&\ket{0_h1_g}_{\text{Alice}}\otimes\ket{0_h1_g}_\text{Bob}\notag\\
\xrightarrow{\text{BS1}}\ket{\psi_{\rm I}}\equiv&
\frac{\ket{0_h1_g}_\text{A}+\ket{1_h0_g}_\text{A}}{\sqrt{2}}\otimes\frac{\ket{0_h1_g}_\text{B}+\ket{1_h0_g}_\text{B}}{\sqrt{2}}
\label{eq1}.
\end{align}

\paragraph{Quantum communication}\ \\
\ \ \ \ \ These entangled states generated in (I) are shared by Alice and Bob after they exchange their guest ports
\begin{align}
\ket{\psi_{\rm I}}=&\frac{1}{2}\Bigl[\ket{1_h0_g}_\text{A}\ket{1_h0_g}_\text{B}+\ket{0_h1_g}_\text{A}\ket{1_h0_g}_\text{B}\notag\\
&\hspace{1.5pc}+\ket{1_h0_g}_\text{A}\ket{0_h1_g}_\text{B}+\ket{0_h1_g}_\text{A}\ket{0_h1_g}_\text{B}\Bigr]\notag\\
\xrightarrow{\text{\tiny Exchanging}}\ket{\psi_{\rm II}}\equiv&
\frac{1}{2}\Bigl[\ket{1_h0_g}_\text{A}\ket{1_h0_g}_\text{B}+\ket{1_h1_g}_\text{A}\ket{0_h0_g}_\text{B}\notag\\
&\hspace{1.5pc}+\ket{0_h0_g}_\text{A}\ket{1_h1_g}_\text{B}+\ket{0_h1_g}_\text{A}\ket{0_h1_g}_\text{B}\Bigr].
\end{align}
\paragraph{Local quantum operation}\ \\
\ \ \ \ \ A $\pi/2$-shift in the guest ports results in
\begin{align}
\hspace{-2pc}\ket{\psi_{\rm III}}
\equiv&\frac{\ket{\Psi^+}_\text{A}\ket{\Psi^-}_\text{B}+\ket{\Psi^-}_\text{A}\ket{\Psi^+}_\text{B}}{2}\notag \\
&\hspace{1.5pc}
+i\frac{\ket{1_h1_g}_\text{A}\ket{0_h0_g}_\text{B}+\ket{0_h0_g}_\text{A}\ket{1_h1_g}_\text{B}}{2}\label{res0},
\end{align}
where $\ket{\Psi^\pm}$ represents the Bell states, which is defined as $\tfrac{1}{\sqrt{2}}\left[\ket{0_h1_g}\pm\ket{1_h0_g}\right]$ and $\ket{1_h1_g}_{\text{A,(B)}}$ represents the existence of two photons in Alice (Bob). 
\paragraph{Measurement}\ \\ \vspace{-1pc}

The $\ket{\psi_\text{III}}$ in Eq.\eqref{res0} indicates that the state is no longer symmetric between Alice and Bob. Then, they obtain different measurement results by distinguishing $\ket{\Psi^\pm}$ and $\ket{1_h1_g}$ states from the others by two-photon interference.
We combine the host and the guest ports with a beam splitter (BS2) as shown in FIG.\ref{cir}. The beam splitters transform the $\ket{\psi_\text{III}}$state into
\begin{align}
\ket{\psi_\text{III}}\xrightarrow{\text{BS2}}&
i\frac{1}{2\sqrt{2}}\bigl[
\overbrace{
\ket{2_\text{A1}0_\text{A2}0_\text{B1}0_\text{B2}}
-\ket{0_\text{A1}2_\text{A2}0_\text{B1}0_\text{B2}}}
^{\Downarrow \ket{1_h1_g}_\text{A}\ket{0_h0_g}_\text{B}}\notag\\
&\hspace{3pc}+
\overbrace{
\ket{0_\text{A1}0_\text{A2}2_\text{B1}0_\text{B2}}
-\ket{0_\text{A1}0_\text{A2}0_\text{B1}2_\text{B2}}}
^{\Downarrow \ket{0_h0_g}_\text{A}\ket{1_h1_g}_\text{B}}
\bigr]\notag \\
&+\frac{1}{2}\Bigl[
\underbrace{\ket{1_\text{A1}0_\text{A2}0_\text{B1}1_\text{B2}}}_{\Uparrow \ket{\Psi^+}_\text{A}\ket{\Psi^-}_\text{B}}
+\underbrace{\ket{0_\text{A1}1_\text{A2}1_\text{B1}0_\text{B2}}}_{\Uparrow \ket{\Psi^-}_\text{A}\ket{\Psi^+}_\text{B}}\Bigr].\label{eq4}
\end{align}
\begin{table}[b]
\begin{center}
\begin{tabular}{ccccclc} \hline\hline
\multicolumn{1}{l}{{\scriptsize Detector}} & \multicolumn{1}{l}{A1} & \multicolumn{1}{l}{A2} & \multicolumn{1}{l}{B1} & \multicolumn{1}{l}{B2} & Leader & \multicolumn{1}{l}{{\scriptsize Probability}} \\ \hline
\multirow{4}{*}{(i)} & 2 & - & - & - & Alice & \multirow{4}{*}{$\dfrac{1}{8}$}\\
 & - & 2 & - & - & Alice & \\
 & - & - & 2 & - & Bob & \\
 & - & - & - & 2 & Bob & \\ \hline
\multirow{2}{*}{(iii)} & 1 & - & - & 1 & Alice & \multirow{2}{*}{$\dfrac{1}{4}$}\\
 & - & 1 & 1 & - & Bob & \\ \hline\hline
\end{tabular}
\caption{Expected results of leader election for all expected combinations of four detectors(A1,A2,B1, and B2):
(i) Either Alice or Bob received two photons. (iii) Alice detects photon at detector A1 and Bob detects it at detector B2 and vice versa.}
\label{table1}\end{center}
\end{table} %
The first four terms of Eq.\eqref{eq4} indicates two-photon detection events at detector A1$\sim$B2, and these terms are those from $\ket{1_h1_g}$.
The second term is from $\ket{\Psi^\pm}_\text{A}\ket{\Psi^\mp}_\text{B}$. Utilizing two-photon interference, $\ket{\Psi^+}_\text{A,B}$ and $\ket{\Psi^-}_\text{A,B}$ states are transformed into one-photon detection on detector A1(B1) and A2(B2), respectively. 
TABLE \ref{table1} lists the expected results for the leader election circuit for all the possible combination of photo-detection events obtained from Eq.\eqref{eq4}. 
According to the TABLE \ref{table1}, 
only one party receives two photons or, single photon at detector 1. 
Therefore, if we define
{\it the party who receives two photons or single photon at detector 1 is the leader}, a unique leader can be elected. 
The present implementation works in principle deterministically even with linear optics.
It should be noted that no classical communication is required during the process.
In TABLE \ref{table1}, Alice and Bob know that if either of them receives two photons, the other receives none. Additionally, if either of them receives one photon at detector "1", the other receives one at "2".
Therefore, Alice recognizes that she has become the leader without classical communication if she receives two photons , or one photon on detector A1. 

In the experiment, we utilize spontaneous parametric downconversion (SPDC) to prepare the photon pairs and single photon detectors to detect them, the efficiencies of which are finite. This implies that experimental QLE protocol cannot be implemented deterministically, in the sense that the photon pairs are generated and detected with certain probability.
Hence, in order to evaluate the performance of QLE experiment, we discuss only the "ideal" condition where the photon pairs are generated and detected, by post-selecting the two photon detection events.
In such ideal condition, QLE protocol works correctly with our implementation with the probability of 1.
Therefore we claim that our implementation is "deterministic" in the rest of this paper.
Moreover, this definition of "deterministic" is commonly used to evaluate the linear optical quantum gates and our implementation is advantageous to the conventional linear optical quantum gates that is "non-deterministic" even with an ideal condition
\footnote{
In the CNOT gates \cite{CNOT1}, in addition to the two photon detection, only the case which one and only one photon is in each output ports must be $post$-$selected$. On the other hand our implementation can operate when the two photon has been emerged in one output.}
.

\section{Experiment}
\begin{figure}[bb]
\includegraphics[width=8.5cm]{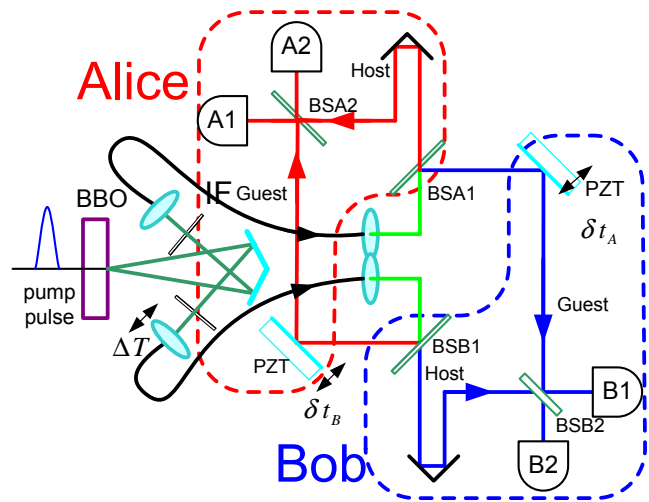}
\caption{(Color online) The experimental setup for two-party QLE. A femtosecond laser pulses (100fs, 404nm, and 82MHz) were used to generate photon pairs by spontaneous parametric down-conversion(SPDC) in 2-mm-thick $\beta$-BBO crystal.}
\label{exp}
\end{figure}
The experimental setup to demonstrate linear optical QLE protocol is shown in FIG.\ref{exp}. 
This setup corresponds to two combined Franson's interferometer\hspace{-0.3pc} \cite{fras}, where one of Alice's arm is exchanged for Bob's counterpart.
As noted before, we used a pair of photons generated by SPDC to prepare initial state $\ket{0_h1_g}_{A}\otimes\ket{0_h1_g}_{B}$. A femtosecond UV pulse laser at 404nm for a duration of 100fs, and a frequency of 82MHz pumped 2-mm-thick $\beta$-BBO crystal, then two down-converted photons were simultaneously generated by SPDC. 
After passing through a bandpass filter of 808 nm with a FWHM of 3 nm, the down-converted photons were coupled into polarization maintaining fiber (PMF) and delivered into twin interferometers. Each photon was divided by beams splitters (BSA1 and BSB1) to provide entangled states $\ket{\psi_\text{I}}$ and then reflected ports ($guest$) were exchanged to share the entanglement. The host and the guest ports were recombined by BSA2 and BSB2, whose the relative phase was controlled by the position of a mirror with a piezo-electric translation (PZT) stage. Finally, we detected photons at A1, A2, B1, and B2 by using single photon counting modules (Perkin-Elmer SPCM), and analyzed the coincidence counts between the detectors. As noted in the end of previous section, we focus only on the two photon detection events, in order to remove the contribution of single photon detection.
The single photon detection is caused not only by the loss of initial photon pair, but by the distinguishable photon pairs\cite{grice} generated with non-degenerated SPDC process, one counterpart of which has been blocked by the bandpass filter. 
In the latter case, one can argue that the correct initial state had not been generated.
The observed coincidence counts are listed in TABLE \ref{table3}.
\begin{table}[t]
\begin{center}
\begin{tabular}{c|cccc}
 & A1 & A2 & B1 & B2 \\ \hline
A1 & (i)$^\prime$3918 & - & - & - \\
A2 & (ii)18\phantom{00} & (i)$^\prime$3896 & - & - \\
B1 & (iv)16\phantom{00} & (iii)137\phantom{0} & (i)$^\prime$3831 & - \\
B2 & (iii)160\phantom{0} & (iv)8\phantom{000} & (ii)11\phantom{00} & (i)$^\prime$3920
\end{tabular}
\caption{Observed coincidence counts of each pattern. Note that diagonal elements are single counts because our detector could not distinguish photon number. Note that only (i)$^\prime$ indicates the single counts.}\label{table3}
\end{center}
\end{table}
Cells (i)$^\prime$ and (ii)$\sim$(iv) in TABLE \ref{table3} represent single counts and coincidence counts, respectively. Ideally, (i)$^\prime$ only includes two-photon detection by one detector. However, the actual counts, (i)$^\prime$, in TABLE \ref{table3} includes the contribution of the single photon detection as mentioned before. 
Since we used threshold detectors in the experiment, we estimate the number of two-photon detection by considering the loss.
Let $\eta \sim 0.02$ be the total transmission of experimental components, the detail of which is shown in TABLE \ref{lossta}. 
\begin{table}[h]
\footnotesize
\begin{tabular}{lc}
\hline\hline
Components&T\\
\hline
Bandpass filter (FWHM3nm)&$\sim$0.1\\
Fiber coupling(single-mode)&$\sim$0.5\\
Beam splitter&$\sim$0.9\\
Fiber coupling(multi-mode)&$\sim$0.8\\
Quantum efficiency&$\sim$0.5\\
Total&	$\sim$0.02\\
\hline\hline
\end{tabular}
\caption{Transmittance of experimental components analysed separately}\label{lossta}
\end{table}
\ \\
The number of single counts (i) that we observed in the experiment can be written as
\begin{align}
N_{single}= 2N_0(1-\eta)\eta+N_0\eta^2,
\end{align}
where $N_0$ indicates the number of generated photon pairs soon after the nonlinear crystal. 
Single counts consist of both one-photon detection, and two-photon detection and the number of two-photon detections can be written as
\begin{align}
N_{2 photon}=N_0\eta^2.
\end{align}\\\
Then, we can estimate the number of two-photon detections $N_{2photon}$ by using the value of $\eta$ measured in a different experiment as
\begin{align}
\frac{N_{2photon}}{N_{single}}=\frac{N_0\eta^2}{2N_0(1-\eta)\eta+N_0\eta^2}\sim\eta/2=0.01\label{eqloss}
\end{align} 
Equation \eqref{eqloss} implies that the actual contribution of two-photon detections in single counts can be 1\%.
\begin{figure}
\begin{center}
\includegraphics[width=6cm]{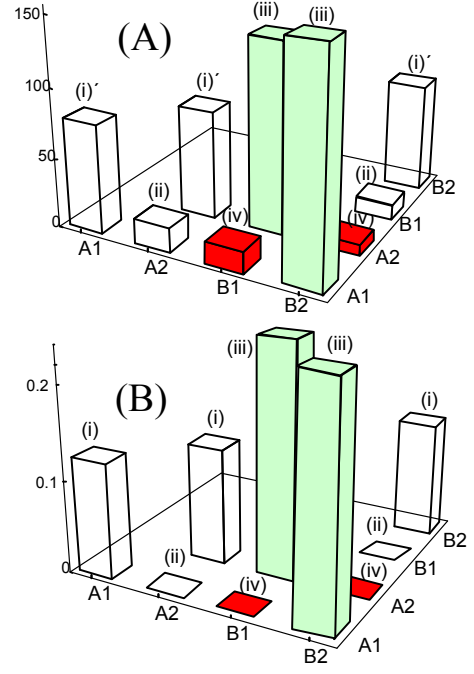}
\caption{(Color online) (A) The histogram of observed coincidence patterns and (B)theoretically predicted values:
(i) two-photon detection on A1, A2, B1, and B2 (corrected),
(ii) coincidence detection A1 AND A2, and B1 AND B2,
(iii) coincidence detection A1 AND B2, and B1 AND A2, and
(iv) coincidence detection A1 AND B1, and A2 AND B2}\label{baro}
\end{center}
\end{figure}
The corrected coincidence counts are plotted in FIG.\ref{baro}(A). The theoretically predicted values from TABLE \ref{table3} have been shown in FIG.\ref{baro}(B) for comparison. Figure\ref{baro}(A) shows good agreement with (B).
Bars (i) and (iii) in FIG.\ref{baro} indicate corrected two-photon detection and $\ket{\Psi^\pm}$ detection, respectively. In these cases leader election is successfully achieved.
Bars (ii) and (iv), on the other hand, result from imperfect two-photon interference.  However, bars (ii) can be regarded as successful events because the detection of Alice and Bob is different. 
Bars (iv) indicates that the same detector (A1\&B1, and A2\&B2) clicked on Alice and Bob, then leader election failed.
Hence, the imperfect two-photon interference is responsible for the performance of the experimental QLE protocol.

Since we focus only on the two photon events, the success probability per trial of QLE circuit can be calculated by:
\begin{align}
P=1-\frac{{\rm N_{(iv)}}}{{\rm N_{(i)}+N_{(ii)}+N_{(iii)}+N_{(iv)}}}\label{eq3},
\end{align}
where ${\rm N}_{\rm (i)}\sim {\rm N}_{\rm (iv)}$ represent the counts of (i)$\sim$(iv). 

\begin{figure}[h]
\begin{center}
\includegraphics[width=6cm]{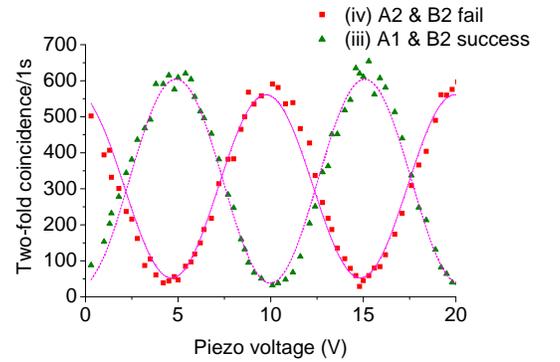}
\caption{(Color online) Measured two-photon interference fringe}\label{fringe}
\end{center}
\end{figure}

To calculate the success probability from Eq.\eqref{eq3}, we must derive the number of two-photon detections N$_{\rm (i)}$. Since the N$_{\rm (i)}$ counts are obtained from rough estimates of losses as previously, we cannot precisely calculate the success probability. Therefore, we rewrite Eq. \eqref{eq3} using the coincidence probabilities between four detectors \eqref{eqap5}, derived in Appendix \ref{app1}, as
\begin{align}
P^\prime&=\frac{4\gamma}{(1+\gamma)^4}\bigl[1+(1+\nu)\gamma+\gamma^2\bigr]\label{eq43}\\
&\hspace{3.2cm} \gamma\equiv R/T \notag
\end{align}
where $\gamma=0.8$ and $\nu$ correspond to the branching ratio of beam splitters and the visibility of two-photon interference.  Using Eq.\eqref{eq43}, we can evaluate performance of QLE circuit with the imperfection of two-photon interference, which is characterized by visibility $\nu$.
To estimate the visibility $\nu$, we observed the two-photon interference fringes shown in FIG.\ref{fringe}. We recorded two patterns of coincidence counts: (iii) A1\&B2, and (iv) A2\&B2, which correspond to success and failure, respectively, as a function of the voltage applied to PZT.
Both coincidence counts fit the sinusoidal curves with visibility $\nu= 0.845\pm0.02$.
Hence, the success probability of leader election is calculated as 95.0$\pm$0.48\%, and the operation is therefore almost deterministic.
\section{Discussion}
Let us examine how advantageous the experimental results were over the classical limit. To do this, we should not consider only success probability but also communication complexity. Communication complexity is characterized by the costs of communications between computers to complete algorithms. It is commonly recognized in computer science that performance of distributed algorithms can be evaluated with the communication complexity.

As shown in Appendix \ref{app2}, the optimal classical algorithm involves repeating and comparing the one-bit random number generated by both Alice and Bob until they generate a different number.
The expected minimum costs of communication $N_{classical}$, to complete leader election is calculated as 4.

The quantum leader election, on the other hand, is completed by operating the quantum circuits shown in FIG. \ref{cir}.
As noted in Sec.II, Alice and Bob can finish leader election without the classical communication.
The total communication costs between Alice and Bob is 2-qubit quantum communication to exchange their guest ports. It is clear that ideal quantum leader election is advantageous over classical algorithms because leader election is completed deterministically only with 2-qubit quantum communication.

However, the two photon interference is not perfect in the experiment.
That causes a failure of QLE with a finite probability (P=95.0$\pm$0.48\% per one trial)
\footnote{
We discarded single photon detection events.
Including the single photon detection events will increase the success probability close to unity, which may be useful for practical applications. However, it should be noted that the single photon detection events are not appropriate to examine the performance of QLE protocol, because those states may not satisfy the condition that Alice and Bob must prepare an identical symmetric initial state
} .
In the circuit depicted by FIG2, the detection pattern for Alice and Bob become the same ((iv) A1\&B1 or A2\&B2) with certain probability.
Then, if Alice receives a single photon at "A1", Bob might receive one at "B1". She cannot determine at which detector Bob has actually received a single photon. Since the leader must be elected uniquely, only the successful detection event (iii) must be {\it post-selected}.
Therefore, Alice and Bob have to inform one another of their detection events using 2-bit classical communication to verify their detection events.
\begin{figure}[t]
\begin{center}
\includegraphics[width=7cm]{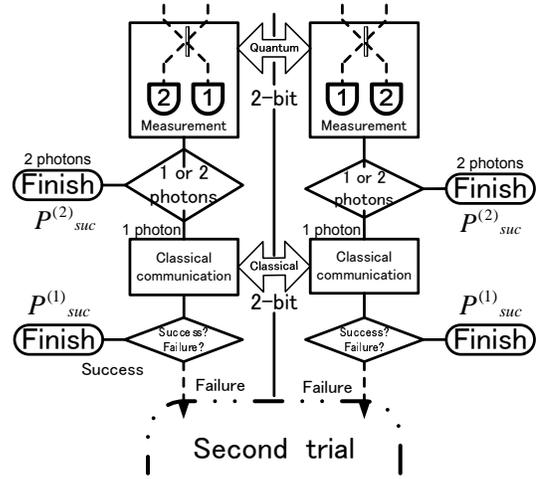}
\caption{
The flow chart for QLE protocol with imperfect interference.}
\label{flow}
\end{center}
\end{figure}
The expected communication costs can be calculated along the flow chart given in FIG.\ref{flow}:
Alice and Bob check the photon number they received. If they received two photons with a probability of $P^{(2)}_{suc}$, then the leader election is completed only with 2-qubit quantum communication. Otherwise, they need to verify the results with 2-bit classical communication.
If their detection event is (iii) with a probability of $P^{(1)}_{suc}$, leader election is completed with a total of 4-bit communication, otherwise it fails with a probability of $1-P^{(2)}_{suc}-P^{(1)}_{suc}(=1-P_{suc})$.
The probability of failure until the $k$-1 th trial is written by $(1-P_{suc})^{k-1}$ and the amount of communication to complete leader election at the $k$-th trial before and after the verification process are written as $4(k-1)+2$ and $4(k-1)+4$.
Then the expected communication costs in QLE protocol $N_{quantum}$ is calculated as:
\begin{align}
N_{quantum}=&\sum_k^\infty 
\Bigl[4(k-1)+2\Bigr]P^{(2)}_{suc}(1-P_{suc})^{k-1}\notag\\
&\hspace{1pc}+\sum_k^\infty\Bigl[4(k-1)+4\Bigr]P^{(1)}_{suc}(1-P_{suc})^{k-1}\label{eqq1}
\end{align}
We calculated Eq.\eqref{eqq1} by substituting Eq.\eqref{eqap3} and \eqref{eqap4} and the $N_{quantum}$ is
\begin{figure}[t]
\begin{center}
\includegraphics[width=8cm]{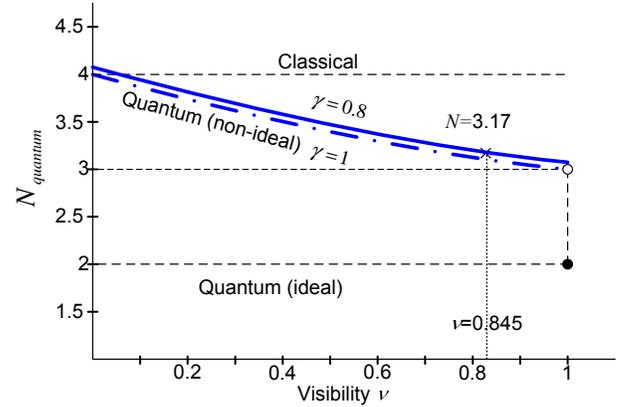}
\end{center}
\caption{
(Color online) Expected communication costs $N_{quantum}$ (in bits and qubits) of quantum leader election as a function of the interference visibility $\nu$ of Alice and Bob's photons.
}\label{complot}
\end{figure}
plotted in FIG.\ref{complot} as a function of two-photon interference fringe visibility $\nu$. Note that without two-photon interference ($\nu$=0) the $N_{quantum}$ is equivalent to the classical limit. However, as shown in FIG.\ref{complot} even if two-photon interference is perfect ($\nu$=1, $\gamma$=1) $N_{quantum}$ does not become that of the ideal case, but approaches to three.
This extra one-bit is still leveraged for the verification of the detection events, because Alice and Bob never know that the operation is in the ideal regime.
We obtain a visibility $\nu$ of 0.845$\pm0.02$\%, the expected total communication costs can be calculated as 3.17$\pm$0.06 which overcomes the classical limit.  

Finally, we touch upon the extension of our implementation for more than three party case.
In the case of multi parties, the QLE protocol requires CNOT operation, which is no longer deterministic with linear optics.
Moreover all the parties must prepare a cat state and exchange them. The probability to generate such multi photon states with SPDC would be much lower and the optical circuit would be much more complicated than two party case.
 For the above reasons, it is difficult to implement multi party QLE protocol with liner optics.
\section{Conclusion}
We demonstrated the experimental 2-party quantum leader election.
Unlike conventional probabilistic quantum circuits based on linear optics, our leader election circuit can operate deterministically only with linear optics. We derived the success probability $P_{suc}$ and expected total communication costs to complete leader election $N_{quantum}$ in terms of two-photon interference visibility in order to evaluate our experimental result.
Finally, we have obtained nearly deterministic results for a success probability of 95.0$\pm$0.48\% and an expected total communication costs of 3.17$\pm$0.06, both of which improve upon the classical limit.
\section*{Acknowledgment}
We would like to thank Dr. K. Matsumoto and Dr. M. Hayashi for their helpful discussions, and especially K. Yoshino for his kind suggestion and advice on the experiments.
\appendix
\section{Calculation of two-photon interference}\label{app1}
We calculate the coincidence probabilities between four detectors by calculating the overlap integral of field operators.
The field operator just before detectors A1, A2, B1, and B2 can be written as follows
\begin{align}
\hat{E}^\dagger_{A1}(t)=&T\hat{E}^\dagger_s(t-t_{Ah})+R\hat{E}^\dagger_i(t-t_{Ag})\label{Ea1}\\
\hat{E}^\dagger_{A2}(t)=&\sqrt{RT}\hat{E}^\dagger_s(t-t_{Ah})-\sqrt{RT}\hat{E}^\dagger_i(t-t_{Ag})\label{Ea2}\\
\hat{E}^\dagger_{B1}(t)=&\sqrt{RT}\hat{E}^\dagger_i(t-t_{Bh})+\sqrt{RT}\hat{E}^\dagger_s(t-t_{Bg})\label{Eb1}\\
\hat{E}^\dagger_{B2}(t)=&T\hat{E}^\dagger_i(t-t_{Bh})-R\hat{E}^\dagger_s(t-t_{Bg})\label{Eb2}
\end{align}
where $R$ and $T$ are the reflectance and the transmittance of four identical beam splitters (BSA1, BSB1, BSA2, and BSB2), $t_{Ah}, t_{Ag}, t_{Bh}, and  t_{Bg}$ are the time delay of each optical paths, and $\hat{E}_s\hat{E}_i$ are the field operators of down-converted photons.
The field operator, $\hat{E}$, can be written by the Fourier transform of the creation operator.
\begin{align}
\hat{E}_{k}(t)\equiv \int d\omega\ a_k^\dagger(\omega) e^{-i\omega t}\label{appfi},
\end{align}
The probability of coincidence detection between detector $i,j$ is given by
\begin{align}
P_{i,j}=\int d\tau \int dt \Bigl|\hat{E}_{i}(t)\hat{E}_{j}(t+\tau)\ket{\psi}\Bigl|^2\label{keisan0}
\end{align}
where $\ket{\psi}$ indicates the wave function of the input state, which is defined \cite{gho} as
\begin{align}
\ket{\psi}=&\int d\omega_p\ \xi(\omega_p,\omega_0)\int d\omega\  \phi(\omega)\notag \\
&\hspace{3pc}\hat{a}^\dagger_\text{s}\bigl(\tfrac{\omega_p}{2}+\omega\bigr)
\hat{a}^\dagger_\text{i}\bigl(\tfrac{\omega_p}{2}-\omega\bigr)\ket{0}
\label{appparame}
\end{align}
where $\omega_p$ is the center frequencies of the pump beam. The spectral broadening of the pump and the down-converted photons are introduced as $\xi_p(\omega_p,\omega_0)$ and $\phi(\omega)$, respectively.
By combining Eqs.\eqref{Ea1}$\sim$\eqref{appparame}, and after a straightforward algebra, we obtain the probabilities of two-fold coincidence and two-photon detection as follows:
\begin{align}
P^{(2)}_{A1}=&P^{(2)}_{A2}=\gamma^2T^4\biggl[1+e^{-\frac{\Delta\omega^2}{2}\delta t_A^2}\biggr]\label{eqf}\\
P^{(2)}_{B1}=&P^{(2)}_{B2}=\gamma^2T^4\biggl[1+e^{-\frac{\Delta\omega^2}{2}\delta t_B^2}\biggr]\label{eqb}\\
P^{coin}_{A1,A2}=&\gamma(1+\gamma^2)T^4\biggl[1-\tfrac{2\gamma}{1+\gamma^2}e^{-\frac{\Delta\omega^2}{2}\delta t_A^2}\biggr]\label{eqa}\\
P^{coin}_{B1,B2}=&\gamma(1+\gamma^2)T^4\biggl[1-\tfrac{2\gamma}{1+\gamma^2}e^{-\frac{\Delta\omega^2}{2}\delta t_B^2}\biggr]\\
P^{coin}_{A1,B1}=&(1+\gamma^4)T^4\biggl[1+\tfrac{2\gamma^2}{1+\gamma^4}
e^{-\frac{\Delta \omega_p^2}{32}(\delta t_A+\delta t_B)^2}\notag\\
&\hspace{2pc}\times\hspace{1pc}e^{-\frac{\Delta\omega^2}{8}(\delta t_A-\delta t_B)^2}\cos\tfrac{\omega_0}{2}(\delta t_A+\delta t_B)\biggr]
\end{align}
\begin{align}
P^{coin}_{A2,B2}=&(\gamma^2+\gamma^2)T^4\biggl[1+\tfrac{2\gamma^2}{\gamma^2+\gamma^2}
e^{-\frac{\Delta \omega_p^2}{32}(\delta t_A+\delta t_B)^2}\notag\\
&\hspace{2pc}\times\hspace{1pc}e^{-\frac{\Delta\omega^2}{8}(\delta t_A-\delta t_B)^2}\cos\tfrac{\omega_0}{2}(\delta t_A+\delta t_B)\biggr]\\
P^{coin}_{A1,B2}=&(\gamma+\gamma^3)T^4\biggl[
1-\tfrac{2\gamma^2}{\gamma+\gamma^3}e^{-\frac{\Delta \omega_p^2}{32}(\delta t_A+\delta t_B)^2}\notag\\
&\hspace{2pc}\times\hspace{1pc}e^{-\frac{\Delta\omega^2}{8}(\delta t_A-\delta t_B)^2}\cos\tfrac{\omega_0}{2}(\delta t_A+\delta t_B)\biggr]\\
P^{coin}_{A2,B1}=&(\gamma+\gamma^3)T^4\biggl[
1-\tfrac{2\gamma^2}{\gamma+\gamma^3}e^{-\frac{\Delta \omega_p^2}{32}(\delta t_A+\delta t_B)^2}\notag\\
&\hspace{2pc}\times\hspace{1pc}e^{-\frac{\Delta\omega^2}{8}(\delta t_A-\delta t_B)^2}\cos\tfrac{\omega_0}{2}(\delta t_A+\delta t_B)\biggr]\label{eql}
\end{align}
To obtain Eq .\eqref{eqf}$\sim$\eqref{eql}, we introduce $\delta t_{A}(\equiv t_{Ah}-t_{Ag}),\delta t_{B}(\equiv t_{Bh}-t_{Bg}), and \gamma(\equiv R/T)$, which represent the time difference of two ports on Alice and Bob, and the branching ratio of beam splitters, respectively.
Here, we have made the following assumptions for simplicity: (1)$\delta t_A=\delta t_B=\delta t$, i.e., the time difference is the same for Alice and Bob, since they are forced to do the same operations during the QLE protocol. 
We also assumed 
(2)$\Delta \omega_p =2\Delta\omega$. i.e., the spectral broadening of the photon pair is half that of pump beam ($\Delta\omega_p=2\Delta\omega$). 
Further, we assumed (3) $\tfrac{\omega_0}{2}(\delta t_A+\delta t_B)$=0, i.e., the relative phase between the $host$ and the $guest$ in Alice and Bob is set to 0. Therefore, the coincidence probabilities can be approximated as
\begin{align}
P^{(2)}_{A1}=P^{(2)}_{A2}=P^{(2)}_{B1}=&P^{(2)}_{B2}=\frac{\gamma^2}{\left(1+\gamma\right)^4}\bigl[1+\nu\bigr],\label{eqap1}\\
P_{A1,A2}=P_{B1,B2}=&\gamma\frac{1+\gamma^2}{\left(1+\gamma\right)^4}\bigl[1-\tfrac{2\gamma}{1+\gamma^2}\nu\bigr],\\
P_{A1,B1}=&\bigl[1-\tfrac{2\gamma^2}{1+\gamma^4}\nu\bigr],\\
P_{A2,B2}=&\frac{2\gamma^2}{\left(1+\gamma\right)^4}\bigl[1-\nu\bigr],\\
P_{A1,B2}=P_{A2,B1}=&\gamma\frac{1+\gamma^2}{\left(1+\gamma\right)^4}\bigl[1+\tfrac{2\gamma^2}{\gamma+\gamma^3}\nu\bigr]\label{eqap2}
\end{align}
where $\nu\equiv e^{-\frac{\delta \omega^2}{2}\delta t^2}$ implies the temporal overlapping of two photon wave packets on beam splitters. Using\eqref{eqap1}$\sim$ \eqref{eqap2} we obtain the probability of two-photon detection $P^{(2)}_{suc}$ and single photon detection $P^{(1)}_{suc}$ as
\begin{align}
P^{(2)}_{suc}=&P^{(2)}_{A1}+P^{(2)}_{A2}+P^{(2)}_{B1}+P^{(2)}_{B2}+P_{A1,A2}+P_{B1,B2}.\notag\\
=&\frac{2\gamma}{(1+\gamma)^2}\label{eqap3}\\
P^{(1)}_{suc}=&P_{A1,B2}+P_{A2,B1}=\frac{2\gamma(1+2\nu\gamma+\gamma^2)}{(1+\gamma)^4}.\label{eqap4}
\end{align}
Hence, the success probability per trial is calculated as
\begin{align}
P_{suc}\equiv& P^{(1)}_{suc}+P^{(2)}_{suc}\notag\\
=&\frac{4\gamma}{(1+\gamma)^4}\bigl[1+(1+\nu)\gamma+\gamma^2\bigr]\ \ \  
\Biggl(\xrightarrow{\gamma,\nu=1}1\Biggr)\label{eqap5}
\end{align}
\section{Optimal Classical Algorithms}\label{app2}
How can we obtain optimal classical algorithms with minimum communication complexity?
We obtain the classical limit, $N_{classical}$, which corresponds to the minimum expected value of the total amount of communication to finish leader election.
The classical algorithm for leader election is to generate and compare the numbers between Alice and Bob, and then the party who generated the larger number becomes the leader. 
The problem is how to generate the numbers and with how many bits to compare them. Let $n$ and $p$ be the number of bits and the probability to generate "1" on each bit( 1-$p$ for "0"), respectively; then, the success probability of leader election per trial $P^{(n)}_{cl}$ is calculated as
\begin{align}
P^{(n)}_{cl}=&1-\sum_{k=0}^{n}
\begin{pmatrix}
\ n\ \\
\ k\ 
\end{pmatrix}
p^{2n}(1-p)^{2(n-k)}
=1-(p^2+(1-p)^2)^n\notag\\
&=1-(2p^2-2p+1)^n.\label{pnp}
\end{align}
Then,the optimal condition is obtained by differentiating Eq.\eqref{pnp} as:
\begin{align}
\pdif{P^{(n)}_{cl}}{p}=n&(2p^2-2p+1)^{n-1}(4p-2)=0\label{deq}
\end{align}
By solving the Eq.\eqref{deq}, the optimal probability, $p$, is obtained as $p=1/2$. The optimal success probability is calculated as
\begin{align}
P^{(n)}_{cl}=1-\frac{1}{2^n}\label{pno}
\end{align}
Hence, the expected value for the total amount of communication, $<N>$, after $n$-bit comparison is written as:
\begin{align}
\langle N \rangle=&\sum_{k=0}^\infty (2n\ k) P^{(n)}_{cl}\biggl(1-P^{(n)}_{cl}\biggr)^{k-1}\notag \\
=&2n\biggl(1-\tfrac{1}{2^n}\biggr)\sum_{k=0}^\infty k\Bigl(\tfrac{1}{2^n}\Bigr)^{k-1}
=\frac{2n}{1-2^{-n}}\geqq 4\label{cls}
\end{align}
Then, the classical limit is calculated as $N_{classical}=4$.

\end{document}